\documentclass[twocolumn,showpacs,preprintnumbers,amsmath,amssymb,superscriptaddress,unsortedaddress]{revtex4} 
\usepackage{graphicx}
\usepackage{dcolumn}
\usepackage{bm}
\usepackage{epsf,epsfig,latexsym}

\newcommand{\Version}{Jan. 21, 2004}

\newcommand{\bnl}           {$\rm^{1}$}
\newcommand{\ires}          {$\rm^{2}$}
\newcommand{\kraknuc}       {$\rm^{3}$}
\newcommand{\krakow}        {$\rm^{4}$}
\newcommand{\baltimore}     {$\rm^{5}$}
\newcommand{\newyork}       {$\rm^{6}$}
\newcommand{\nbi}           {$\rm^{7}$}
\newcommand{\texas}         {$\rm^{8}$}
\newcommand{\bergen}        {$\rm^{9}$}
\newcommand{\bucharest}     {$\rm^{10}$}
\newcommand{\kansas}        {$\rm^{11}$}
\newcommand{\oslo}          {$\rm^{12}$}

\begin{document}

\title{Centrality dependence of charged-particle 
pseudorapidity distributions  from  d+Au
collisions at $\sqrt{s_{NN}}=200~\rm GeV$} 

\author{
  I.~Arsene\bucharest,
  I.~G.~Bearden\nbi,
  D.~Beavis\bnl,
  C.~Besliu\bucharest,   
  B.~Budick\newyork,
  H.~B{\o}ggild\nbi, 
  C.~Chasman\bnl,
  C.~H.~Christensen\nbi, 
  P.~Christiansen\nbi, 
  J.~Cibor\kraknuc, 
  R.~Debbe\bnl, 
  E. Enger\oslo,
  J.~J.~Gaardh{\o}je\nbi,
  M.~Germinario\nbi,  
  K.~Hagel\texas, 
  H.~Ito\bnl$^,$\kansas, 
  A.~Jipa\bucharest, 
  J.~I.~J{\o}rdre\bergen, 
  F.~Jundt\ires, 
  C.~E.~J{\o}rgensen\nbi, 
  R.~Karabowicz\krakow, 
  E.~J.~Kim\bnl$^,$\kansas, 
  T.~Kozik\krakow, 
  T.~M.~Larsen\oslo, 
  J.~H.~Lee\bnl, 
  Y.~K.~Lee\baltimore, 
  S.~Lindal\oslo, 
  R.~Lystad\bergen, 
  G.~L{\o}vh{\o}iden\oslo,
  Z.~Majka\krakow, 
  A.~Makeev\texas, 
  M.~Mikelsen\oslo, 
  M.~Murray\texas$^,$\kansas, 
  J.~Natowitz\texas, 
  B.~Neumann\kansas,
  B.~S.~Nielsen\nbi, 
  D.~Ouerdane\nbi, 
  R.~P\l aneta\krakow, 
  F.~Rami\ires, 
  C.~Ristea\bucharest,
  O.~Ristea\bucharest,
  D.~R{\"o}hrich\bergen, 
  B.~H.~Samset\oslo, 
  D.~Sandberg\nbi, 
  S.~J.~Sanders\kansas, 
  R.~A.~Sheetz\bnl, 
  P.~Staszel\krakow$^,$\nbi, 
  T.~S.~Tveter\oslo, 
  F.~Videb{\ae}k\bnl, 
  R.~Wada\texas, 
  Z. Yin\bergen,
  I.~S.~Zgura\bucharest\\
  (BRAHMS Collaboration )\\[1ex]
  \bnl~Brookhaven National Laboratory, Upton,New York 11973,
  \ires~Institut de Recherches Subatomiques and Universit{\'e} Louis
  Pasteur, Strasbourg, France,
  \kraknuc~Institute of Nuclear Physics, Krakow, Poland,
  \krakow~Jagiellonian University, Krakow, Poland,
  \baltimore~Johns Hopkins University, Baltimore, Maryland 21218,
  \newyork~New York University, New York, New York 10003,
  \nbi~Niels Bohr Institute, University of Copenhagen, Denmark,
  \texas~Texas A$\&$M University, College Station,Texas 77843,
  \bergen~University of Bergen, Department of Physics, Bergen,Norway,
  \bucharest~University of Bucharest,Romania,
  \kansas~University of Kansas, Lawrence, Kansas 66045,
  \oslo~University of Oslo, Department of Physics, Oslo,Norway}
  \noaffiliation

\date{\Version}
\begin{abstract}
Charged-particle pseudorapidity densities are presented for the d+Au reaction
at $\sqrt{s_{NN}}=200\rm~GeV$ with $-4.2 \le \eta \le 4.2$. The results, 
from the BRAHMS experiment at RHIC,
are shown for minimum-bias events and 0-30\%, 30-60\%, 
and 60-80\% centrality classes.    
Models incorporating both soft physics and hard, perturbative
QCD-based scattering physics
agree well with the experimental results. 
The data do not support predictions based on 
strong-coupling, semi-classical QCD.  
In the deuteron-fragmentation region the central
200 GeV data show behavior similar to
full-overlap d+Au results at 
$\sqrt{s_{NN}}=19.4\rm~GeV$. 
\end{abstract}
\pacs{25.75.Dw} 
\maketitle

The saturation of initial parton densities in relativistic
heavy-ion collisions, a manifestation of high-density QCD, is 
expected to significantly influence the pseudorapidity and
centrality dependence of the emitted charged-particle densities 
from these reactions
~\cite{partonsat83,Eskola00,kharzeev01,kharzeev01b,kharzeev03}. 
Earlier charged-particle pseudorapidity density distributions for Au+Au 
collisions~\cite{back00,Star-mult130,adcox01,bearden01a,bearden02,Phobos-mult200-1}
from the Relativistic Heavy-Ion Collider (RHIC) have been
used to constrain model predictions
for ultrarelativistic heavy-ion collisions. They have been 
inconclusive, however, as to whether parton saturation 
in the initial state contributes significantly to the reaction
dynamics, with both saturation-model~\cite{kharzeev01,kharzeev01b}
and calculations that instead
focus on the energy-loss mechanisms for the multiple mini-jets created in
the collisions~\cite{wang91,zhang01,lin01a,lin01b} successfully 
describing the data. 
A similar model ambiguity found in 
explaining the observed suppression
of high-p$_{t}$ particles in Au+Au collisions was recently 
resolved with mid-rapidity 
d+Au data showing the suppression is not an initial-state
effect~\cite{back03,adler03,adams03,arsene03}.
It has been suggested~\cite{kharzeev03}  
that global particle yields in 
d+Au collisions might result in a definitive signature of parton
saturation. 

We report on a measurement of the charged-particle pseudorapidity
densities for the d+Au reaction at $\sqrt{s_{NN}}=200~\rm GeV$ 
with pseudorapidity $\eta$ coverage of $-4.2 \le \eta \le 4.2$. 
The pseudorapidity
densities are reported for minimum-bias events and  0-30\%, 30-60\%,
and 60-80\%
centrality classes. The results allow for a detailed comparison to
model predictions of particle production at RHIC energies. 
The most central data (0-30\%), 
where both deuteron nucleons are expected to
participate in the reaction,
are compared with full-overlap  d+Au data 
obtained by the NA35 collaboration at 
$\sqrt{s_{NN}}=19.4~\rm GeV$~\cite{na35}.

The present analysis employs several of the
BRAHMS global detector subsystems: 
The Si Multiplicity Array (SiMA) and the scintillator Tile
Multiplicity Array (TMA)~\cite{lee04} are 
used for centrality determination and to
measure the pseudorapidity densities close to mid-rapidity. The Beam-Beam
Counter (BBC) arrays are used to reconstruct the collision vertex and 
to determine the
pseudorapidity densities at larger pseudorapidities. 
The ``Inelasticity Counters'' (INEL), developed for the pp2pp
experiment~\cite{pp2pp}, are used for a 
close-to-minimum-bias experiment trigger and to provide
vertex position information in cases where the 
beam-beam counter arrays are not able to establish this information. 
Full details of the BRAHMS apparatus can be found in ref.~\cite{adamczyk}.

The layout of the SiMA and TMA detectors for the d+Au experiment
is similar to that presented for earlier measurements of Au+Au 
multiplicities at
$\sqrt{s_{NN}}=130~\rm GeV$ and 200~GeV, 
and details of the analysis procedures 
can be found in refs.~\cite{bearden01a,bearden02}. 
The SiMA was configured with 25, 
$4~\rm cm \times 6~\rm cm$  
Si wafers in an hexagonal arrangement around the beam 
pipe, with
each wafer functionally divided into 7 discrete segments along the beam
line and 
located 5.3~cm from the 
beam axis.  Four sides of the hexagonal
array were populated with six detectors, each, with 
the remaining two sides left largely unpopulated except for
a single wafer mounted outside the acceptance of either of the BRAHMS
spectrometers~\cite{adamczyk}. 
The TMA was populated with 38, 
$12~\rm cm \times 12~\rm cm$
plastic scintillator tiles with fibre-optic readout located 13.7~cm from the
beam axis. The hexagonal TMA array had four sides fully populated with 
eight detectors, each, with two and four detectors
mounted on the other two sides, respectively. 
With this arrangement,
the SiMA and TMA can each cover the pseudorapidity range 
$-2.2 \le \eta \le 2.2$ for collisions at array center. In the analysis,  
a range of collision vertex locations z 
about the nominal array center is used, with $-15~\rm cm \le z \le 15~\rm cm$. 
Particle 
multiplicities were deduced for an individual SiMA and TMA element
by using GEANT~\cite{Geant} simulations to convert the 
observed energy-loss signal 
to the number 
of primary particles hitting that element.
The HIJING event simulator~\cite{wang91} 
was used to obtain the initial distribution of particle
types and momenta.   

Two Beam-Beam Counter arrays, positioned around the beam pipe on either
side of the nominal interaction point at a distance of 2.20~m,
are used to extend 
coverage out to $\eta = \pm 4.2$. Each array consists of
separate sets of of small (19 mm diameter) and large (51 mm diameter)
Cherenkov UV-transmitting plastic radiators coupled to photomultiplier 
tubes. Leading particles timing achieves a vertex position
resolution of $\approx 2~\rm cm$.   Charged-particle 
multiplicities are deduced from the number of particles hitting
each detector, as found by dividing the measured ADC signal by that
corresponding to a single incident particle. 

Three pairs of INEL Counters
were used to develop a near-to-minimum bias trigger
by detecting charged particles in the pseudorapidity range
$3.2 < |\eta| < 5.3$. 
The basic INEL counter consists of a plastic
scintillator ring that is segmented into four pieces
and arranged
about the beam pipe. 
The counter  locations with respect to the nominal vertex and 
(inner radius; outer radius) were $\pm 155~\rm cm$ 
(4.13~cm; 12.7~cm), $\pm 416~\rm cm$ (6.67~cm; 12.7~cm) 
and $\pm 660~\rm cm$(6.67~cm; 12.7~cm). 
Relative time-of-flight measurements with the INEL arrays lead to
an interaction vertex determination with a resolution of
$\approx 5~\rm cm$. The INEL counters also 
provide a minimum-bias trigger for the 
experiment. Based on GEANT simulations, they are sensitive
to $91\pm3\%$ of the total inelastic cross section. 

Reaction centrality is determined using 
a geometry-weighted average of  SiMA and TMA multiplicities. 
Both the SiMA and TMA multiplicities are corrected for
the distance of the actual interaction vertex from the nominal vertex
at array center.  GEANT simulations were used to 
correct for the possibility that  neither the SiMA nor the TMA detectors will
be hit by a particle for the most peripheral events.  

\begin{figure}
  \epsfig{file=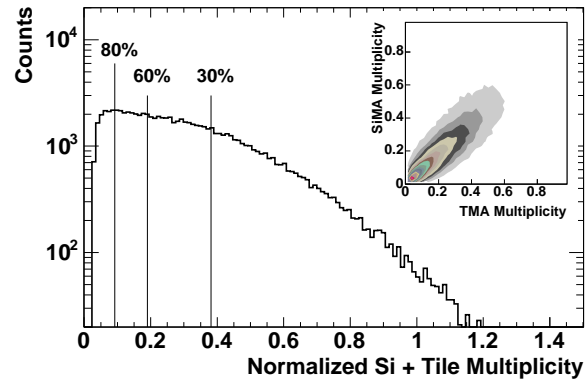,width=8.2cm}
  \caption{
    SiMA and TMA averaged multiplicity distribution normalized to
the 1\% centrality level. 
Lines show efficiency corrected limits for indicated centralities.  
The insert
shows the correlation between the SiMA and TMA multiplicities.}
  \label{multiplicity}
\end{figure}

Figure~\ref{multiplicity} shows the normalized SiMA and TMA
averaged multiplicities. The observed falloff 
is unlike that of the corresponding 
Au+Au spectrum\cite{bearden01a}, where
there is an extended ``flat'' region and  a well-defined high
multiplicity knee.  The d+Au spectrum instead
reflects a smaller number of
participants, making the measurements more
sensitive to the underlying 
nucleon-nucleon collision multiplicity distribution.
Also, the small particle multiplicities for d+Au collisions result in a
relatively large range in the fraction of particles detected
for a given total number of particles emitted.  
The broad correlation band found in comparing SiMA and TMA multiplicities
(Fig.~\ref{multiplicity} insert) 
illustrates the statistical scatter of the semi-independent multiplicity
measurements.  Reaction
centralities are found by integrating the yield under the multiplicity curve.
Limits for the 30\%,  60\%, and 80\% centrality cuts
are indicated by the vertical lines.   

\begin{figure}
  \epsfig{file=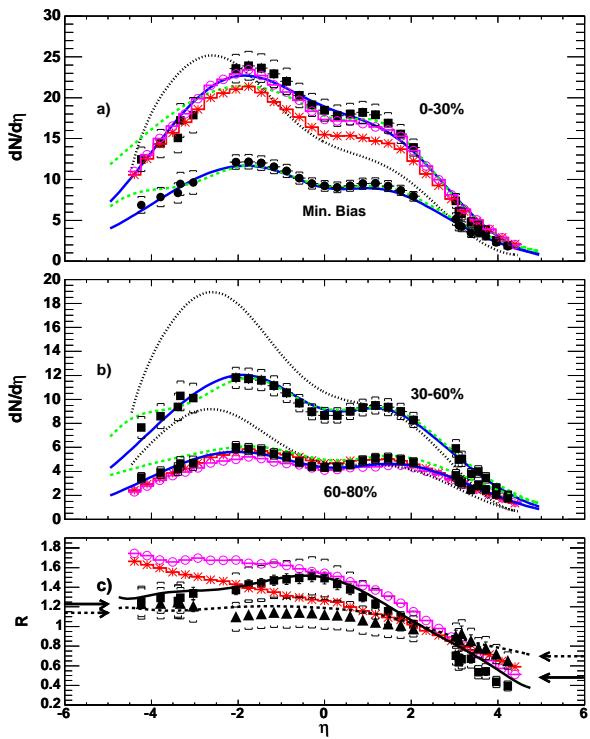,width=8.3cm}
  \caption{
    a) and b) Charged-particle pseudorapidity densities
for indicated centrality ranges. c) Multiplicity
ratios R$^{0-30}$ (squares) and R$^{30-60}$ (triangles), 
as discussed in the text. 
Statistical uncertainties are indicated
by vertical lines or are smaller than the symbols. Detached 
horizontal brackets indicate the total (statistical and systematic)
uncertainties.  
The solid, dashed  and dotted
curves in a) and b) are the results of the  HIJING, AMPT and 
Saturation models, respectively. The curves in c) show the 
HIJING results for R$^{0-30}$ (solid) 
and R$^{30-60}$ (dashed), with the
arrows indicating the values expected for Au- and d-participant, only,
scaling. In all panels, the connected open circles (asterisks) 
correspond to unrestricted HIJING calculations with
centrality classes based on multiplicity (impact paramter),
as discussed in the text.  
 }
  \label{dndeta}
\end{figure}

Figure~\ref{dndeta} shows the resulting charged-particle 
pseudorapidity-density 
plots for minimum-bias events and
0-30\%, 30-60\%, and 60-80\% centrality classes. The SiMA and TMA
results have been averaged. Overall statistical uncertainties are
indicated or are smaller than the data points.  
Systematic  uncertainties, denoted by the horizontal brackets and 
estimated as 8\% for the averaged SiMA and TMA results and 12\% for the
BBC values,
are determined by exploring the variation of
the deduced pseudorapidity densities to reasonable changes in the
energy calibrations and background subtraction.  
Our minimum-bias data agree
within systematic uncertainties with recently reported 
results~\cite{Phobos03}.   

Three model calculations
are compared to the data.  The solid curves show the
predictions of HIJING~\cite{wang91}, a Monte
Carlo model that includes both soft and hard,
perturbative QCD-based scattering effects. The dashed curves
show the predictions of the  
AMPT model~\cite{zhang01,lin01a,lin01b} which includes both
initial partonic and final hadronic interactions.   
For comparison with experiment, both the HIJING and AMPT model 
results have been filtered through a GEANT~\cite{Geant} 
simulation of the BRAHMS
experimental response. Centrality is based on the fraction
of events with the highest particle multiplicity within the
pseudorapidity range of the SiMA and TMA arrays. 
Both models reproduce the experimental
results at midrapidity and at positive rapidities approaching the
deuteron fragmentation region. At negative rapidity (Au 
fragmentation region) the two models start to diverge and
here HIJING appears to be in slightly better agreement with our results. 

The dotted curves in Fig.~\ref{dndeta} show the expectations of the
Saturation Model~\cite{kharzeev03} which accounts for
the high-density QCD effects that are expected to limit the number
of partons in the entrance channel.  In this case the centrality
dependence was based on the published curves of charged-particle
pseudorapidity densities for different centrality ranges
given in ref.~\cite{kharzeev03}. 
The model appears to
be unsuccessful in reproducing either the    
centrality or pseudorapidity dependence of the present results.

The number of participants $N_{part}$ scaled ratios of central-to-peripheral
[R$^{0-30}=(0.35\pm0.03)\times\frac{dN^{0-30\%}}{d\eta}
/\frac{dN^{60-80\%}}{d\eta}$] and
mid-central-to-peripheral
[R$^{30-60}=(0.56\pm0.04)\times\frac{dN^{30-60\%}}{d\eta}/\frac{dN^{60-80\%}}
{d\eta}$] 
charged
particle densities are shown in Fig.~\ref{dndeta}c along with the
corresponding HIJING ratios (curves). Here we take 
$\langle \rm N_{part}\rangle =13.6\pm 0.3$, 
8.5$\pm$0.3, and 4.7$\pm$0.3 for the 0-30\%, 30-60\%, and 
60-80\% centrality ranges, respectively. 
The systematic uncertainties for the ratios include the
participant scaling uncertainty and a 5\% uncertainty for the
experimental pseudorapidity density ratios.
The participant ratios appropriate for Au(left arrows)- and d(right arrows)-
participant-only scaling are shown in Fig 2c for R$^{0-30}$ 
(solid) and
R$^{30-60}$(dashed), respectively.  The HIJING model reproduces well the
experimental ratios, as shown in Fig.~\ref{dndeta}c. In this regard,
it can be noted that the midrapidity pseudorapidity densities 
obtained in a stand-alone HIJING calculation
scale roughly as the number of Au participants.

We use HIJING/GEANT to 
explore the potential bias introduced by the limited
acceptance of the MA (SiMA and TMA) on the deduced pseudorapidity
distributions and R  values. The connected open circles in Fig.~\ref{dndeta}
show the results for the 0-30\% (panel a) and 60-80\% (panel b)
centrality classes
of an unrestricted HIJING calculation where the 
centrality is based on all charged particles emitted in the reaction,
and not just those that satisfy the experimental acceptance.
The greatest effect on the
pseudorapidity distributions is found for the 60-80\% centrality
cut and amounts to as much as an 18\% 
enhancement in the measured dN/d$\eta$ values in the
Au fragmentation region. The corresponding R$^{0-30}$
curve no longer shows evidence of the mid-rapidity maximum observed
for the experimental results. 

In lighter systems, the events selected in a given range of
multiplicity-based centrality are not all the same as would be 
selected if it were possible to base centrality on the impact parameter.  
This is shown in Fig.~\ref{dndeta} where the connected asterisks 
indicate the HIJING model
$dN_{ch}/d\eta$ distributions  for 0-30\% (panel a) and 60-80\% (panel b) 
centrality classes based on impact parameter, and the 
corresponding R$^{0-30}$ ratio (panel c).  
Here the  R$^{0-30}$ curve shows a steady rise from the d- to Au-
fragmentation sides, illustrating that the centrality selection does
affect the deduced $1/N_{part}\times dN_{ch}/d\eta$ values. 
 
The d+Au system was previously studied by the NA35 experiment at
$\sqrt{s_{NN}}=19.4\rm~GeV$~\cite{na35} 
where data  were obtained for both negative hadrons 
$h^-$and for the net baryons as measured by the difference of 
proton and antiproton 
yields ($p-\bar p$). Pseudorapidity distributions were found corresponding to
the most central  43\% of the total inelastic cross section,
with both deuteron nucleons acting as participants.
For comparison with the present results, 
the total charged-particle densities are determined for the
lower energy 
data taking $dN_{ch}/d\eta =2\times dN(h^-)/d\eta + dN(p-\bar p)/d\eta$.
The pseudorapidity densities are deduced from quoted rapidity
distributions by first shifting to the center-of-mass system and then
assuming the $\pi^-$ mass for $h^-$ 
and the observed mean-$p_t$ 
values for the $h^-$ and $p-\bar p$ distributions.  
At the higher energy of the
current measurement, HIJING model simulations indicate that
the criteria that both deuteron nucleons act as participants 
is well satisfied for the 0-30\% centrality range. In this
case it is interesting to see if the limiting fragmentation behavior
previously observed in Au+Au yields at 
$\sqrt{s_{NN}}=\rm 130~and~200~GeV$~\cite{bearden01a,bearden02}
is present in the d+Au system.

\begin{figure}
  \epsfig{file=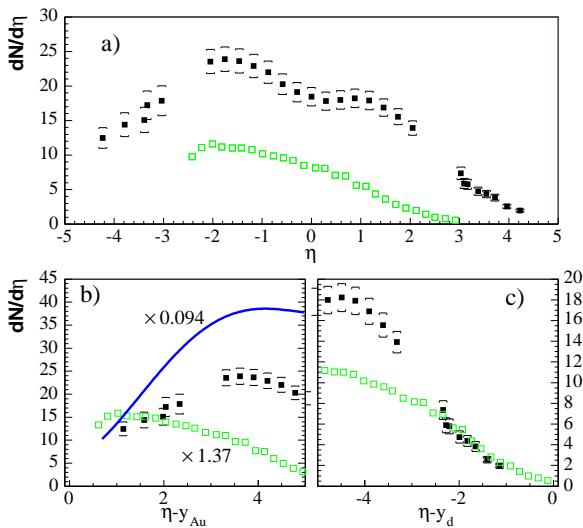,width=8.2cm}
  \caption{
Comparison of central  
$\sqrt{s_{NN}}=200\rm~GeV$ results(solid squares)
with NA35 data (open squares) 
at $\sqrt{s_{NN}}=19.4\rm~GeV$ in a) the nucleon-nucleon 
center-of-mass system, b) the Au rest frame,
c) the deuteron rest frame. The solid curve is based on 
data for Au+Au 0-30\% central events at 
$\sqrt{s_{NN}}=200\rm~GeV$~\cite{bearden02}.
$N_{part}$ scaling is applied as indicated in panel b).
}
\label{na35}
\end{figure}

Figure~\ref{na35}a compares the d+Au pseudorapidity distributions in the
nucleon-nucleon center-of-mass system, where the fixed-target
NA35 results have been shifted by the center-of-mass rapidity. 
A factor of 2.2 increase is seen in the
charged-particle density at mid-rapidity for the higher energy data.
Using HIJING to determine $N_{part}$, 
$\frac{\langle N_{part}^{d+Au}(19.4{\rm~GeV})\rangle}{
\langle N_{part}^{d+Au}({\rm 200~GeV})\rangle}\times\frac{dN^{\rm 200~GeV}/d\eta}
{dN^{\rm 19.4~GeV}/d\eta}=1.7$. Although the different methods of event
selection for the two experiments might reduce this value, the data do not
support simple participant scaling with energy.
Fig.~\ref{na35}b compares the data at the two energies in the frame of
the Au fragment, with the NA35 results scaled up by
the ratio of Au participants at the two energies.  
The two distributions have similar values approaching the
Au rapidity, although it should be noted that
counting the number of participants in the heavier
fragmentation region of a very mass-asymmetric 
reaction is difficult because of
multiple scatterings in the target spectator matter~\cite{na35}.   
The solid curve shows the results for the 0-30\% central
Au+Au distribution at $\sqrt{s_{NN}}=\rm 200~GeV$~\cite{bearden02}, scaled by
the ratio of d+Au gold participants to the number of Au+Au 
participant pairs.  The current measurements
do not extend close enough to the beam rapidities to assess limiting
fragmentation scaling on the Au fragmentation side.    
Fig.~\ref{na35}c compares the two d+Au distributions in the
deuteron frame.  With the given centrality selections,
$N_{part}(\rm d)\approx 2$
at both energies and so no participant scaling is done for the
comparison.  The
two distributions are found to overlap from roughly one to two units of
pseudorapidity below beam rapidity, suggesting a limiting
fragmentation behavior over at least this range. 

Pseudorapidity densities of charged particles for the d+Au
reaction at $\sqrt{s_{NN}}=200~\rm GeV$ are presented 
for different centrality ranges.   
The ratio of particle densities for central and peripheral events is
found to agree well with participant scaling in terms of the respective
fragments away from mid-rapidity.   
Overall, model
calculations based on both soft physics and 
perturbative QCD  (HIJING, AMPT) lead to excellent agreement with the
experimental results.  Calculations based on the saturation
picture using scale parameters set by previous experimental data
fail to reproduce the measurements and lead to a pseudorapidity 
dependence very different from that observed with the current data. 
Comparison with lower energy d+Au data suggests
a limiting fragmentation-like behavior 
near the rapidity of the deuteron fragment. 

We thank the RHIC collider team for their efforts. 
This work was supported by the Office of Nuclear Physics
of the U.S. Department of Energy,  
the Danish Natural Science Research Council, the Research Council of
Norway, the Polish State Committee for Scientific Research (KBN) 
and the Romanian
Ministry of Research.

\end{document}